\documentclass{moriond}

\def\be{\begin{equation}}
\def\ee{\end{equation}}
\def\bea{\begin{eqnarray}}
\def\eea{\end{eqnarray}}

\newcommand{\Photo}{\includegraphics[height=35mm]{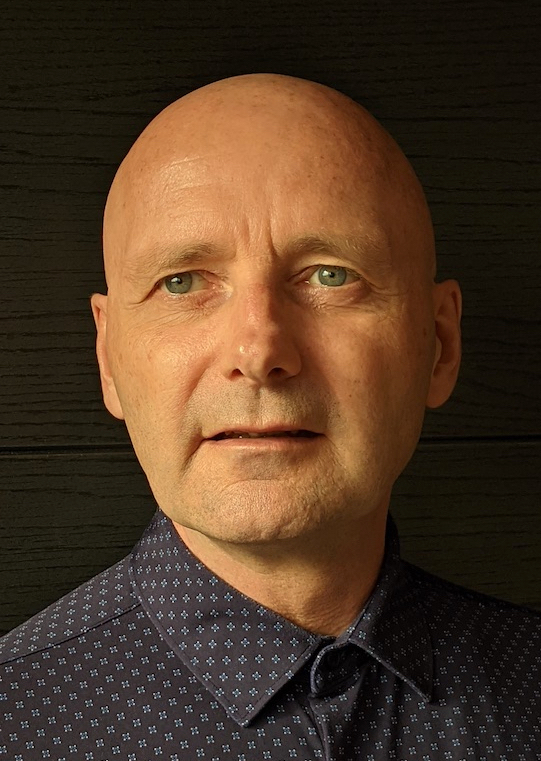}}
\usepackage{amsmath,amssymb}

\begin{document}
\vspace*{4cm}

\title{GLOBAL VISION OF PRECISION MEASUREMENTS}
\author{JENS ERLER}
\address{PRISMA$^+$ Cluster of Excellence, Institute for Nuclear Physics, \\
Johannes Gutenberg-University, 55099 Mainz, Germany\vspace*{12pt}}

\maketitle\abstracts{
I summarize recent developments in electroweak precision physics and global fits.
Expectations for future measurements, both at lower energies and the energy frontier, are also discussed.}

\section{Introduction}
The electroweak (EW) precision program started about half a century ago, and it was very successful in that it predicted
the masses of the $W$, $Z$, and Higgs bosons, and also of the top quark, well before their respective discoveries.
In 2012 the Standard Model (SM) was completed with the discovery of the Higgs boson, and it is as successful as it is unsatisfactory,
in that Dark Matter finds no explanation and there are the problems of naturalness and arbitrariness.
Since no new states have been discovered so far at the LHC, it is possible that history repeats itself,
and that new physics shows up in EW precision measurements first.

As a general remark, the precision in this program has become very high, and the higher it gets, the more physics issues enter
in the interpretation of any given precision measurement.
This can be an obstacle when one is looking at only one observable at a time, but it may become a feature in global analyses,
where one looks across different observables and different subfields of particle, nuclear and atomic physics at the same time,
and one persons uncertainty induced by some subfield of physics may be another persons physics target. 
Right now there are some tensions in the anomalous magnetic moment of the muon, the $W$ boson mass, $M_W$,
and the unitarity of the first row of the CKM matrix, but by and large the SM is still in excellent shape.

\section{Key observables}
\begin{figure}[t]
\begin{minipage}{0.49\linewidth}
\centerline{\includegraphics[width=1.03\linewidth]{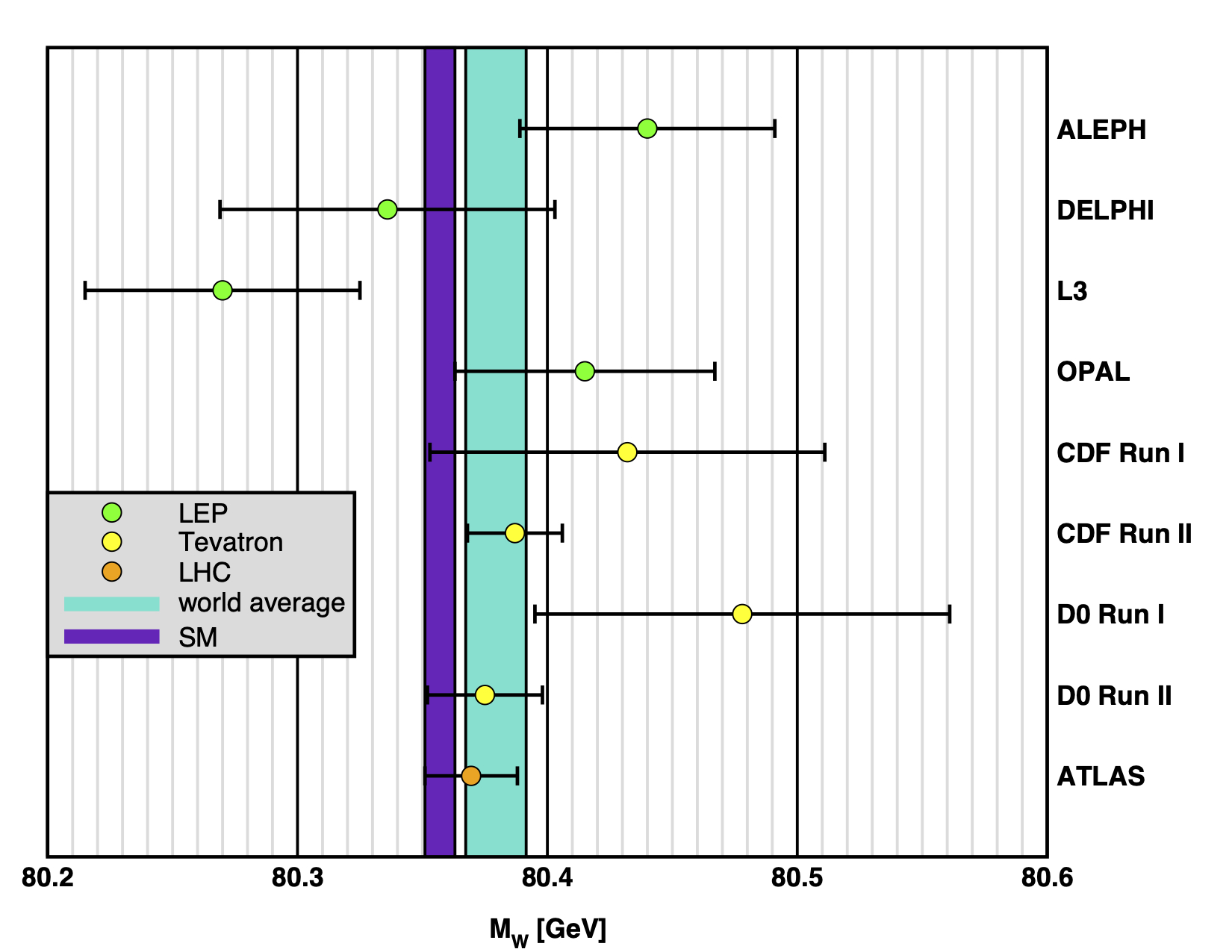}}
\end{minipage}
\hfill
\begin{minipage}{0.49\linewidth}
\centerline{\includegraphics[width=1.03\linewidth]{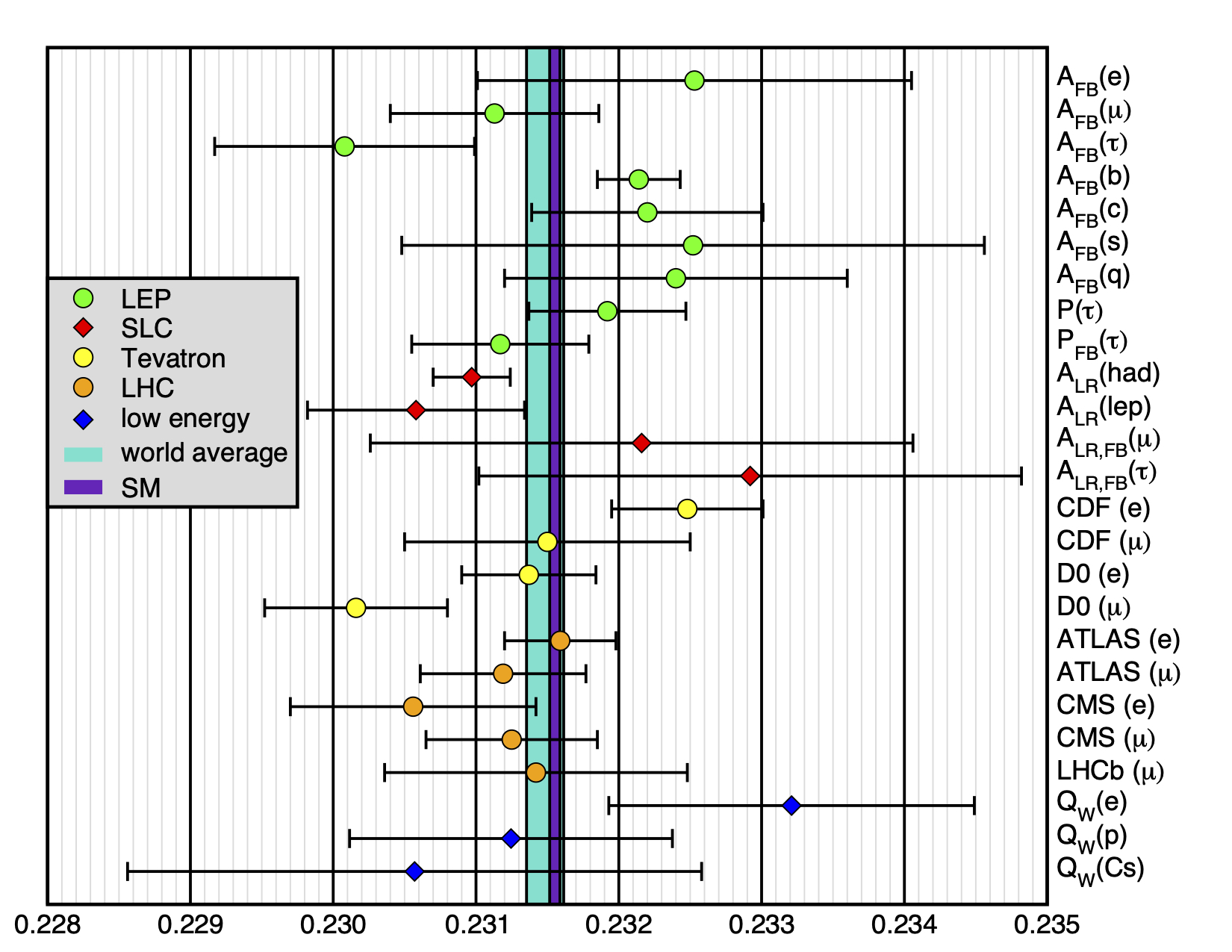}}
\end{minipage}
\caption[]{Left: Measurements of $M_W$ from LEP, the Tevatron and the ATLAS Collaboration at the LHC.
Right: Measurements with sub-percent precision of the effective leptonic weak mixing angle, $\sin^2\theta_{\rm eff}^\ell$.
The world average and the SM prediction are also shown in each plot.}
\label{fig:mw}
\end{figure}

As can be seen from the left plot of Fig.~\ref{fig:mw}, the $M_W$ measurements from different colliders and run periods are
in very good agreement with each other.
Their average, $M_W = 80.379 \pm 0.012$~GeV, however, deviates at the 2~$\sigma$ level from the SM prediction, $M_W = 80.357 \pm 0.006$~GeV.
From the right plot of Fig.~\ref{fig:mw} one observes that the situation is somewhat reversed in the case of weak mixing angle measurements,
where the individual determinations are quite scattered, while the common average, $\sin^2\theta_{\rm eff}^\ell = 0.23148 \pm 0.00013$,
is in perfect agreement with the value, $\sin^2\theta_{\rm eff}^\ell = 0.23153 \pm 0.00004$, preferred by the SM.
The most precise measurements are from the forward-backward asymmetry for $b$ quark final states at LEP, and the left-right asymmetry from the SLC, 
while the determinations from lower energies shown in blue near the bottom of the plot currently show larger uncertainties.
However, there are upcoming experiments at low momentum transfers, such as the MOLLER experiment~\cite{Benesch:2014bas} 
in polarized $e^-$ scattering, and the P2 experiment in polarized elastic $e^-p$ scattering~\cite{Becker:2018ggl}, 
both of which aiming to measure $\sin^2\theta_W$ with a precision comparable to that at LEP and the SLC, as I will discuss in Section~\ref{P2}.
 
\begin{figure}[t]
\begin{minipage}{0.50\linewidth}
\centerline{\hspace{-48pt}\includegraphics[width=1.15\linewidth]{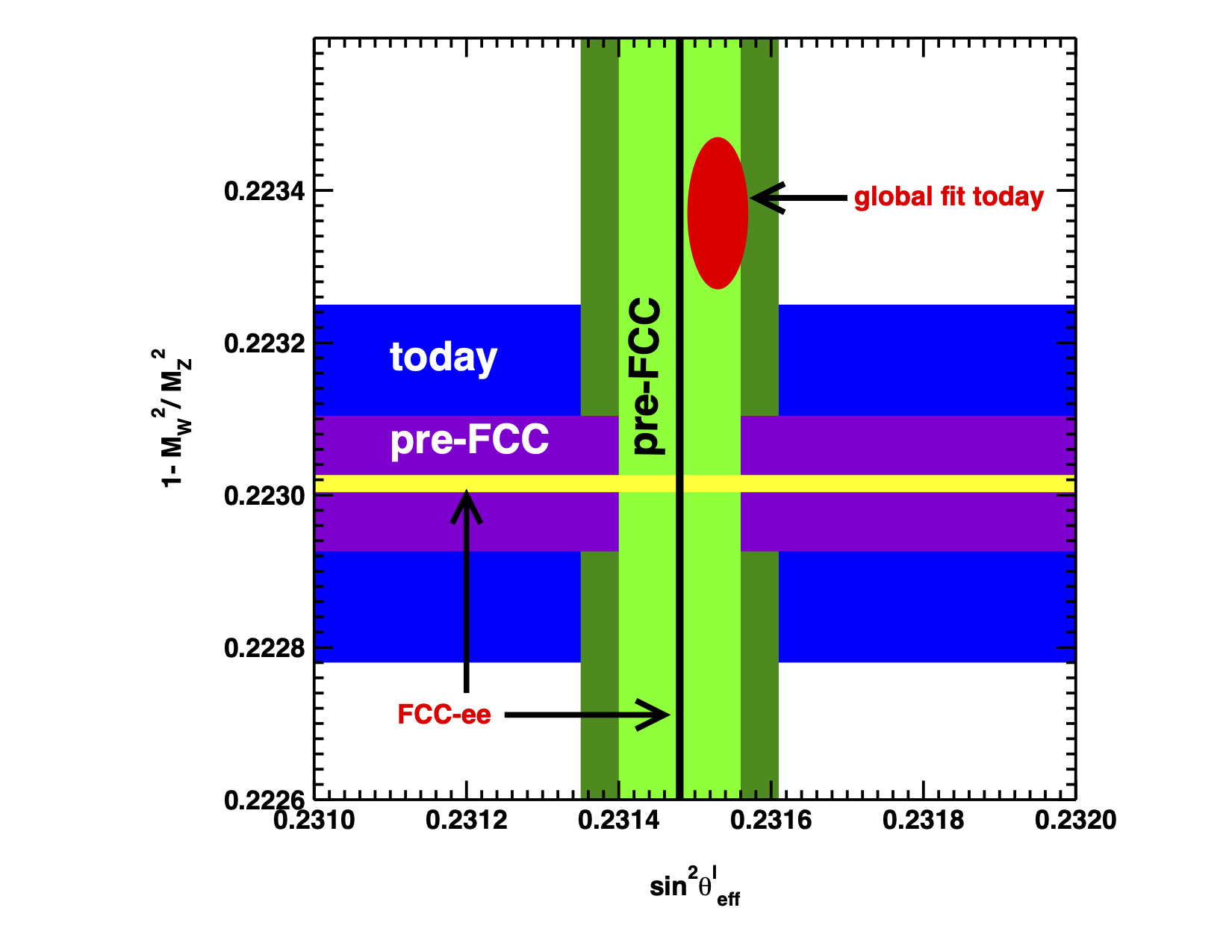}}
\end{minipage}
\hfill
\begin{minipage}{0.48\linewidth}
\centerline{\hspace{-18pt}\includegraphics[width=1.15\linewidth]{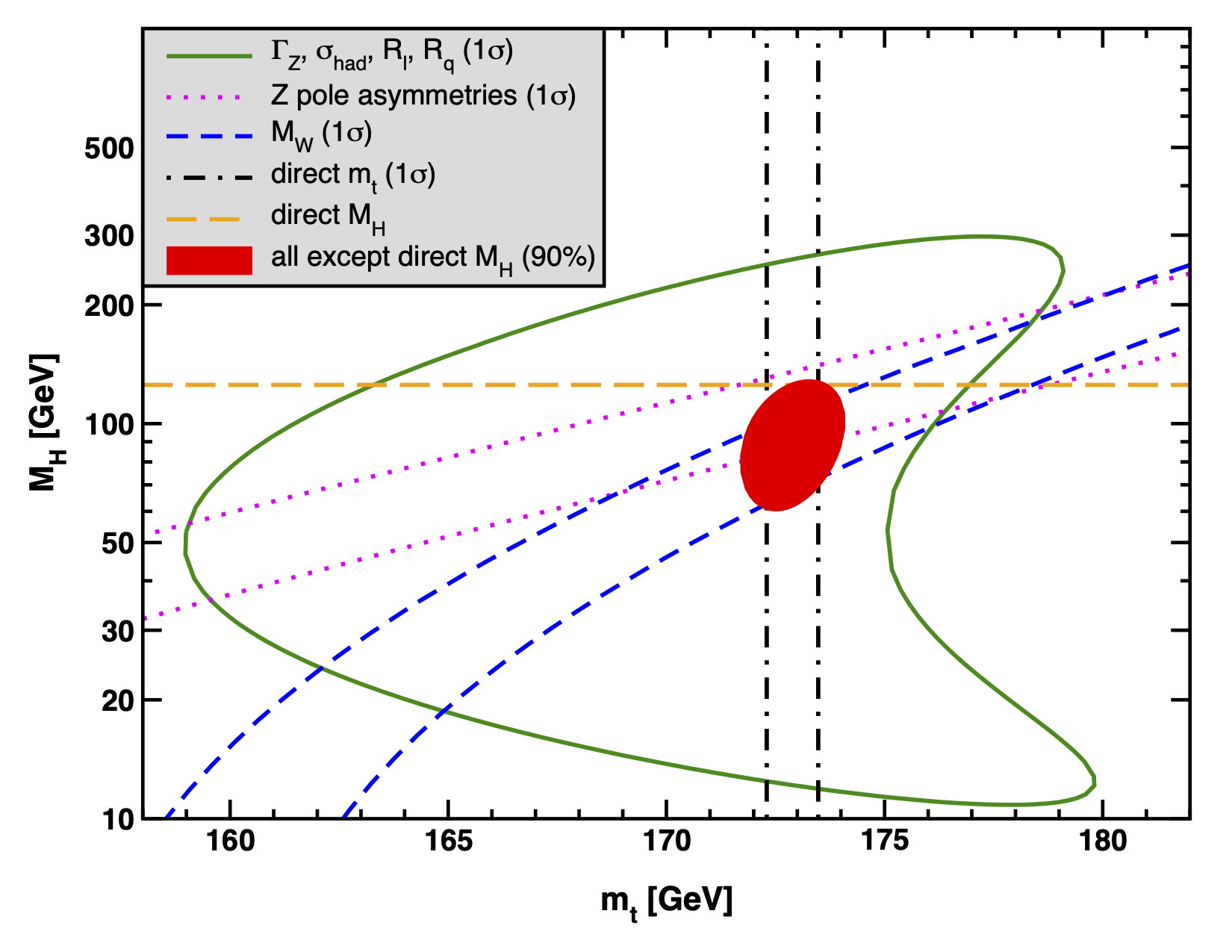}}
\end{minipage}
\caption[]{Left: On-shell {\em vs.}\hspace{-1pt} effective leptonic weak mixing angle.
The wide bands represent the situation today, while the thin lines in the center are the projections of what a future lepton collider
such as the FCC-ee might be able to achieve.
The medium-wide bands indicate what we expect before the start of a new lepton collider, {\em i.e.},\hspace{-1pt} perhaps
within the 15 years ahead of us, namely from further measurements at the LHC but also from low-energy experiments.
The red ellipse represents the current global fit result.
Right: Indirect constraints~\cite{PDG} on $M_H$ and $m_t$ from various $Z$ lineshape observables, $Z$ pole asymmetries, and $M_W$.
The direct determinations are also shown.}
\label{fig:mwfuture}
\end{figure}

The left plot of Fig.~\ref{fig:mwfuture} compares two different definitions of the weak mixing angle and how well they can be measured.
One is the effective leptonic weak mixing angle, $\sin^2\theta_{\rm eff}^\ell$, as determined mostly from cross section asymmetries,
while the vertical axis hosts the on-shell weak mixing angle, given in terms of the ratio of the $W$ and $Z$ boson masses. 
Since these quantities coincide at the tree level, they may be properly normalized with respect to each other,  
and comparing their values one observes that the radiative corrections are quite large, 
and very interesting as they arise from all SM particles and perhaps hitherto unknown ones.
The right plot of Fig.~\ref{fig:mwfuture} summarizes the current situation in terms of the Higgs boson and top quark masses, $M_H$ and $m_t$,
including the direct measurements as the horizontal line and the vertical band, respectively. 
One observes that the asymmetries are in very good agreement with these direct measurements, but that $M_W$ prefers lower values of $M_H$. 
One can remove the kinematic $m_t$ determination from the hadron colliders, and determine~\cite{PDG},
\begin{equation} 
m_t = 176.4 \pm 1.9 \mbox{ GeV}, 
\end{equation} 
which is 1.9~$\sigma$ higher than the average~\cite{PDG} of the direct measurements, $m_t = 172.9 \pm 0.6 \mbox{ GeV}$.

\begin{figure}[t]
\begin{minipage}{0.49\linewidth}
\centerline{\includegraphics[width=1.03\linewidth]{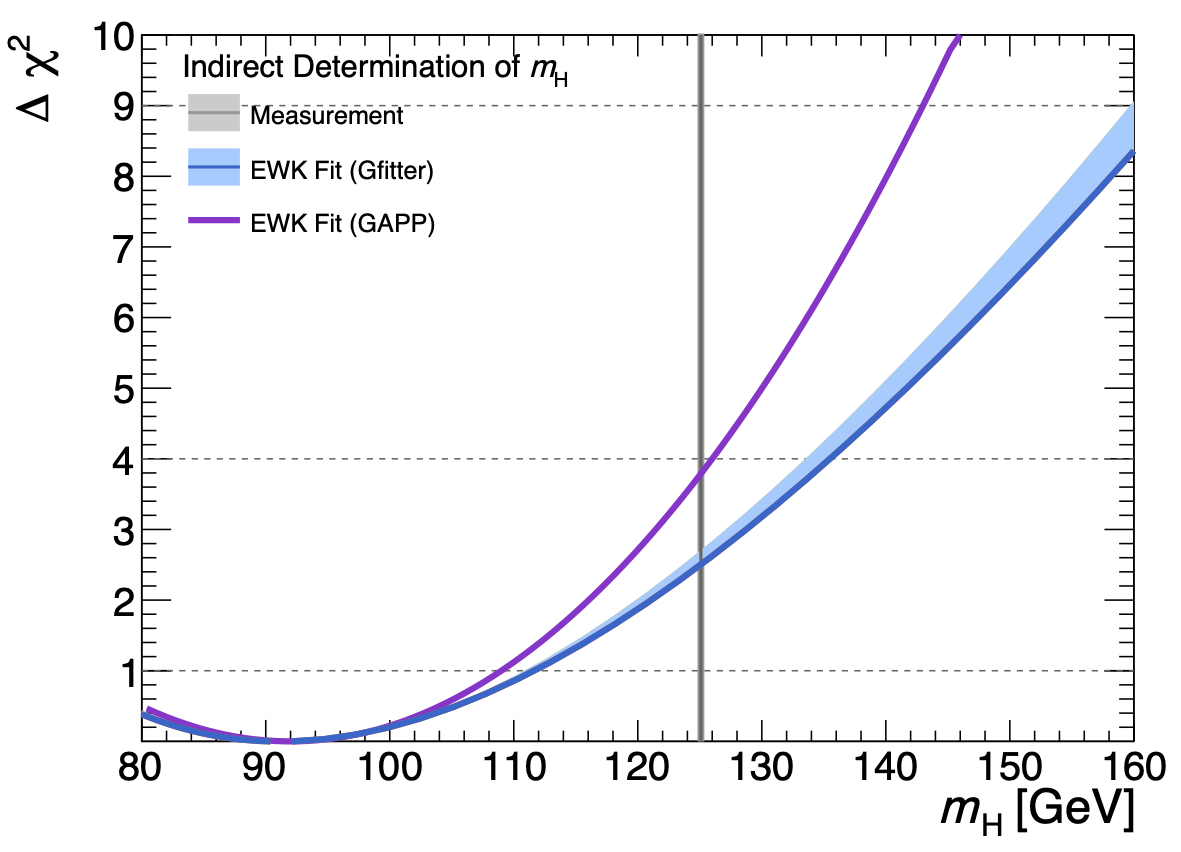}}
\end{minipage}
\hfill
\begin{minipage}{0.49\linewidth}
\centerline{\includegraphics[width=1.03\linewidth]{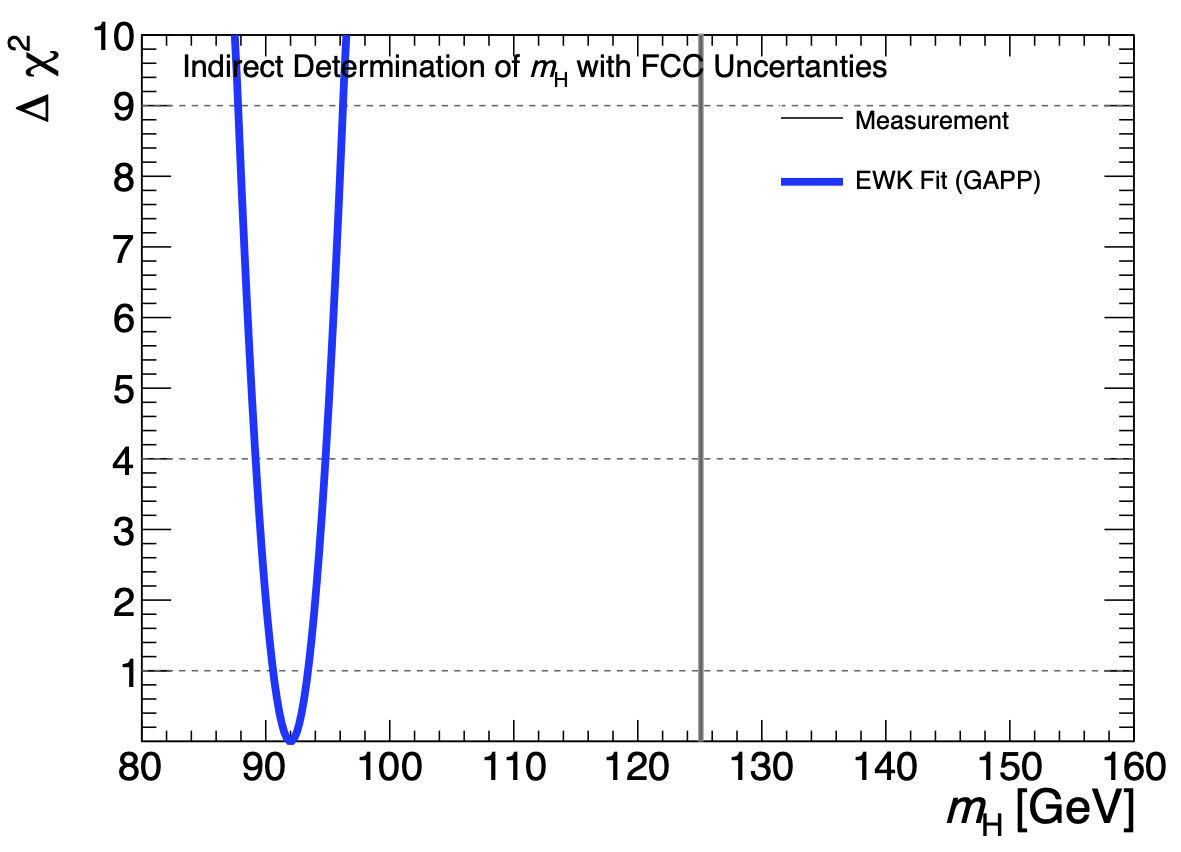}}
\end{minipage}
\caption[]{Left: $\Delta\chi^2 \equiv \chi^2 - \chi^2_{\rm min}$ 
computed~\cite{Erler:2019hds} with the EW packages Gfitter and GAPP as a function of $M_H$.
A modest tension with the kinematic $M_H$ determination at the LHC is visible. 
Right: Same as on the left, but for a fit~\cite{Blondel:2019vdq} assuming the projected FCC-ee errors.}
\label{fig:mhfuture}
\end{figure}

Similarly, if one removes the direct measurements of $M_H$ from Higgs decays at the LHC from the global fit it returns somewhat lower $M_H$.
As can be seen from the left plot of Fig.~\ref{fig:mhfuture} this is only a $\lesssim 2~\sigma$ effect, 
and again a consequence of the measured $M_W$.
The right plot of Fig.~\ref{fig:mhfuture} shows the impact of the projected FCC-ee precision, assuming that the central values of the observables 
would stay where they are today.
In that case, the uncertainty in the indirect $M_H$ would be only $\pm 1.4$~GeV, and the discrepancy of many standard deviations would
allow to claim the discovery of new physics.
This assumes, however, the absence of any theory uncertainties.
Assuming instead the opposite (and unrealistic) extreme of no theory improvement between now and the conclusion of such a lepton collider,
the uncertainty in the indirect $M_H$ would be $\pm 5.7$~GeV.
This would still signal a significant discrepancy, but it would be just shy of a discovery. 

\section{Parity-violating electron scattering}
\label{P2}
\begin{figure}[t]
\begin{minipage}{0.57\linewidth}
\centerline{\vspace{4pt}\hspace{-0pt}\includegraphics[width=1.09\linewidth]{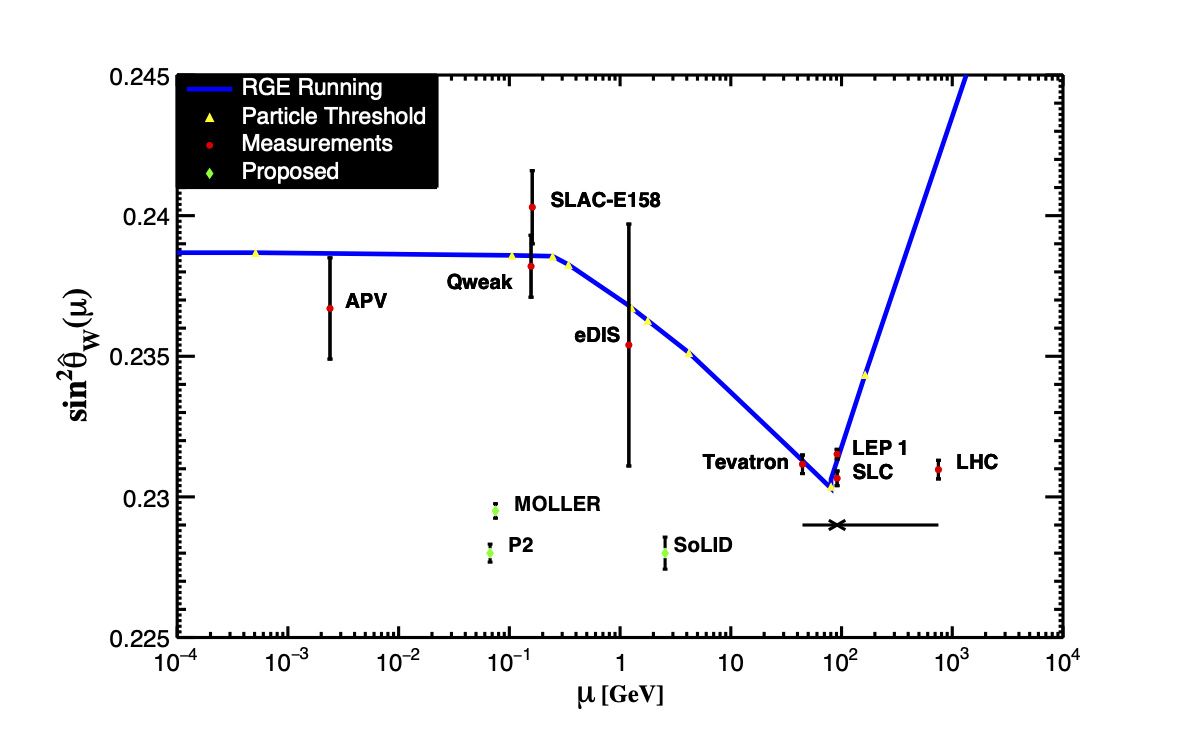}}
\end{minipage}
\hfill
\begin{minipage}{0.42\linewidth}
\centerline{\hspace{-10pt}\includegraphics[width=1.09\linewidth]{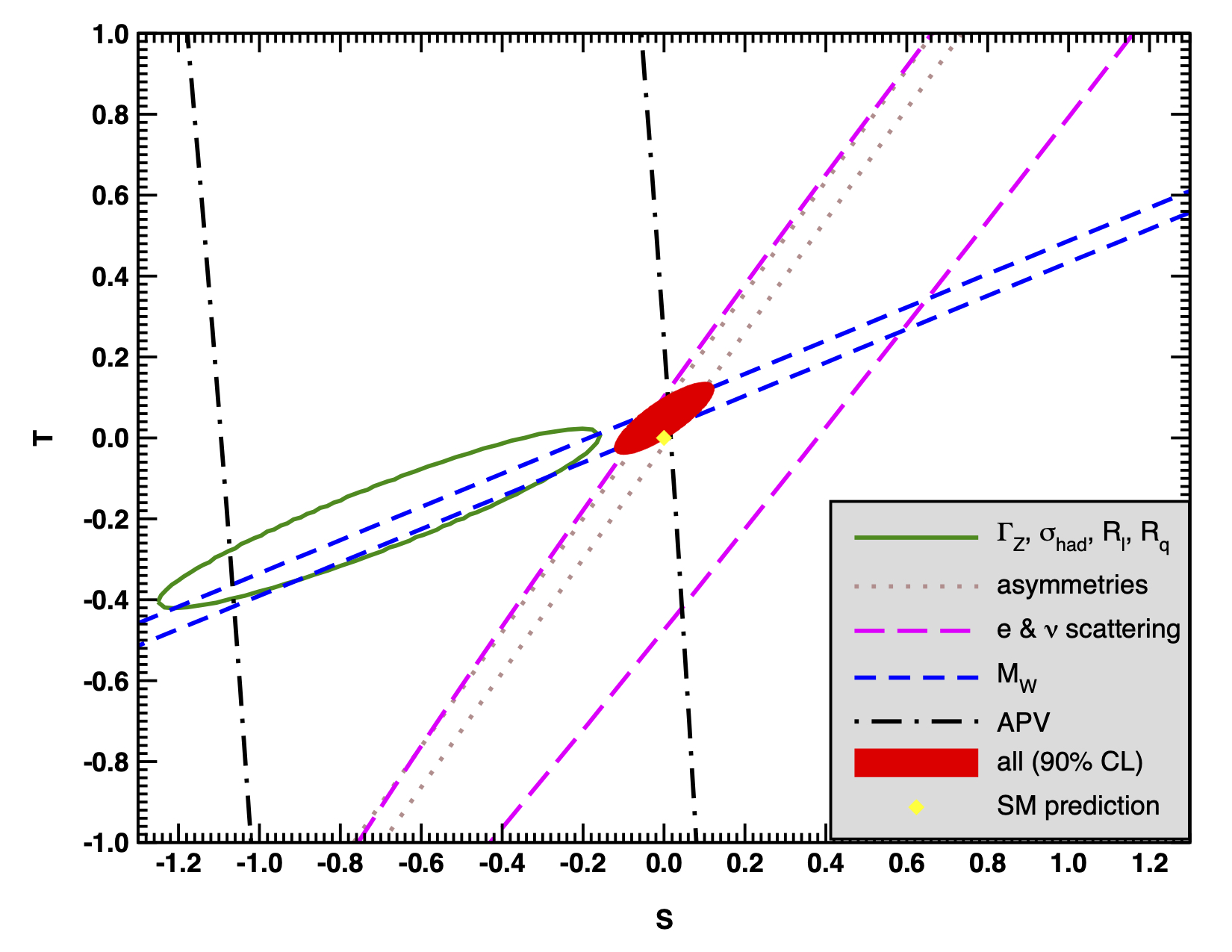}}
\end{minipage}
\caption[]{Left: Scale dependence of the weak mixing angle in the $\overline{\rm MS}$ renormalization scheme~\cite{Erler:2017knj} 
together with the measurements.
Notice, that the associated $\beta$-function changes sign near $M_W$, where the theory effectively switches from non-Abelian to Abelian in character. 
The lower part shows expectations from the next-generation experiments, MOLLER, P2, and SoLID, 
building on E--158 ($e^-$ PVES at SLAC), Qweak, and PVDIS (JLab), respectively.
Right: Current constraints~\cite{PDG} on the oblique $S$ and $T$ parameters.
Further oblique parameters, such as $U$, have been set to zero, as they enter the SMEFT not before the $D=8$ level.}
\label{fig:weakplot}
\end{figure}

The left plot of Fig.~\ref{fig:weakplot} shows that there is more to the weak mixing angle than just averaging different measurements, 
as they may derive from different energy regimes.
Most of them have been obtained from a window around the $Z$ boson mass, including the currently most precise ones from LEP, the SLC,
the Tevatron, and the LHC.
There is also a smaller number of determinations at lower energies, such as from atomic parity violation (APV), $\nu$ scattering, 
and polarized $e^-$ scattering.
Thus, one can test the running of $\sin^2\theta_W$ in the SM, and if there is a new degree of freedom somewhere below the $Z$ scale,
say with a mass of the order of 100~MeV such as a dark $Z$ boson~\cite{Davoudiasl:2012qa}, 
then the running would change and the lower energy measurements would move together away from the predicted curve.
It is also possible that there is some other kind of new effect, such as additional amplitudes, in which case 
only some of the low-energy measurements would tend to deviate, and one might be able to deduce the presence of new physics in this way.

In parity-violating electron scattering (PVES) experiments one directs an $e^-$ beam onto a fixed target, 
and measures the difference in cross sections for left-handed and right-handed polarized beams normalized to the sum.
In proton scattering one needs to work with very low momentum transfers, $|Q^2| \ll m_p^2$, 
in order to scatter elastically from a proton as a whole and to keep hadronic uncertainties under control.  
Such asymmetries filter out the EW interaction as QED and QCD conserve parity.
The challenge is that they are proportional to $G_F Q^2$, with $G_F$ the Fermi constant, and therefore tiny. 
To leading order, PVES asymmetries arise from the interference of photon and $Z$ boson exchange amplitudes,
and since for both, $e^-$ and $p$ scattering, it is proportional to $1 - 4 \sin^2\theta_W$
(similar to leptonic asymmetries near the $Z$ pole) this results in an enhanced sensitivity to $\sin^2\theta_W$.
Using this strategy, the Qweak Collaboration~\cite{Androic:2018kni} was able to extract,
\be
\sin^2\theta_W(0) = 0.2383 \pm 0.0011,
\ee
{\em i.e.},\hspace{-1pt} with 0.5\% precision, from a 4.1\% asymmetry measurement at an average $|Q| = 158$~MeV.
In the future this kind of low-energy experiment will allow per mille level $\sin^2\theta_W$ determinations.

Assuming that there are no new particles with masses near or below the EW scale,
one can describe physics Beyond the Standard Model (BSM) in a completely model-independent way in terms of the SM effective field theory (SMEFT). 
This is because the SM is really just the collection of the operators above the double line in Table~\ref{tab:SMEFT} 
which are the interactions with the nice property of being renormalizable,
but there is no known physics principle which would imply a restriction to operators of dimension $D \leq 4$.
Indeed, the operators at the $D=5$ level account for neutrino oscillations, and these days one is looking in particular at $D=6$ operators.

\begin{table}[b!]
\caption[]{Classification~\cite{Henning:2015alf} of SMEFT operators with $D \leq 8$.
To calculate the numbers of independent operators, partial integrations, algebraic identities, and the equations of motion have been used,
but not re-phasings or other field re-definitions. 
Hermitian conjugate operators are counted separately so that their coefficients may be taken as real parameters.
Kinetic terms have not been counted and topological terms may be misidentified.
The first column is for $N_F = 3$ families of fermions, while all other columns assume $N_F = 1$ for simplicity.}
\label{tab:SMEFT}
\vspace{0.4cm}
\begin{center}
\begin{tabular}{|c||c||c|cccc||c|}
\hline
& $N_f = 3$ & $N_f = 1$ & bosonic & $\psi^2$ & $\psi^4~(\Delta B = 0)$ & $\psi^4~(\Delta B \neq 0)$ & effect \\ \hline\hline
$D = 0$ & 1 & 1 & 1 & --- & --- & --- & $\Lambda_C \neq 0$ \\ \hline
$D = 1$ & --- & --- & --- & --- & --- & --- & \\ \hline
$D = 2$ & 1 & 1 & 1 & --- & --- & --- & $M_H \neq 0$ \\ \hline
$D = 3$ & --- & --- & --- & --- & --- & --- & \\ \hline
$D = 4$ & 55 & 7 & 1 & 6 & --- & --- & \\ \hline\hline
$D = 5$ & 12 & 2 & --- & 2 & --- & --- & $m_\nu \neq 0$ \\ \hline
$D = 6$ & 3045 & 84 & 15 & 31 & 30 & 8 & \\ \hline
$D = 7$ & 1542 & 30 & --- & 10 & 12 & 8 & \\ \hline
$D = 8$ & 44807 & 993 & 89 & 386 & 420 & 98 & \\ \hline
\end{tabular} 
\begin{tabular}{l} \\ \\ \\ \\ \\ \vspace{-10pt}
SM \\ \\ \\ \\ \\ \vspace{-6pt}
BSM 
\end{tabular} 
\end{center}
\end{table}

Let us now focus on the 38 types of four-fermion operators ($N_F = 1$).
They may be categorized as purely leptonic ($3~L^4$), semi-leptonic ($13~L^2 Q^2$), baryon number-violating ($8~L Q^3$),
and purely hadronic ($14~Q^4$), where the latter are difficult to constrain with precision due to strong interaction uncertainties.
Abbreviating, $\psi_V \equiv \bar\psi\gamma^\mu \psi$ and $\psi_A \equiv \bar\psi\gamma^\mu \gamma^5 \psi$,
the three leptonic ones break further down into $e_V e_V$, $e_A e_A$, and $e_V e_A$, where the first two have been studied at LEP and the SLC,
while the last one is parity-odd and can be directly probed by MOLLER (and its precursor experiment~\cite{Anthony:2005pm} E--158 at SLAC). 

\begin{figure}[t]
\begin{minipage}{0.48\linewidth}
\centerline{\includegraphics[width=1.0\linewidth]{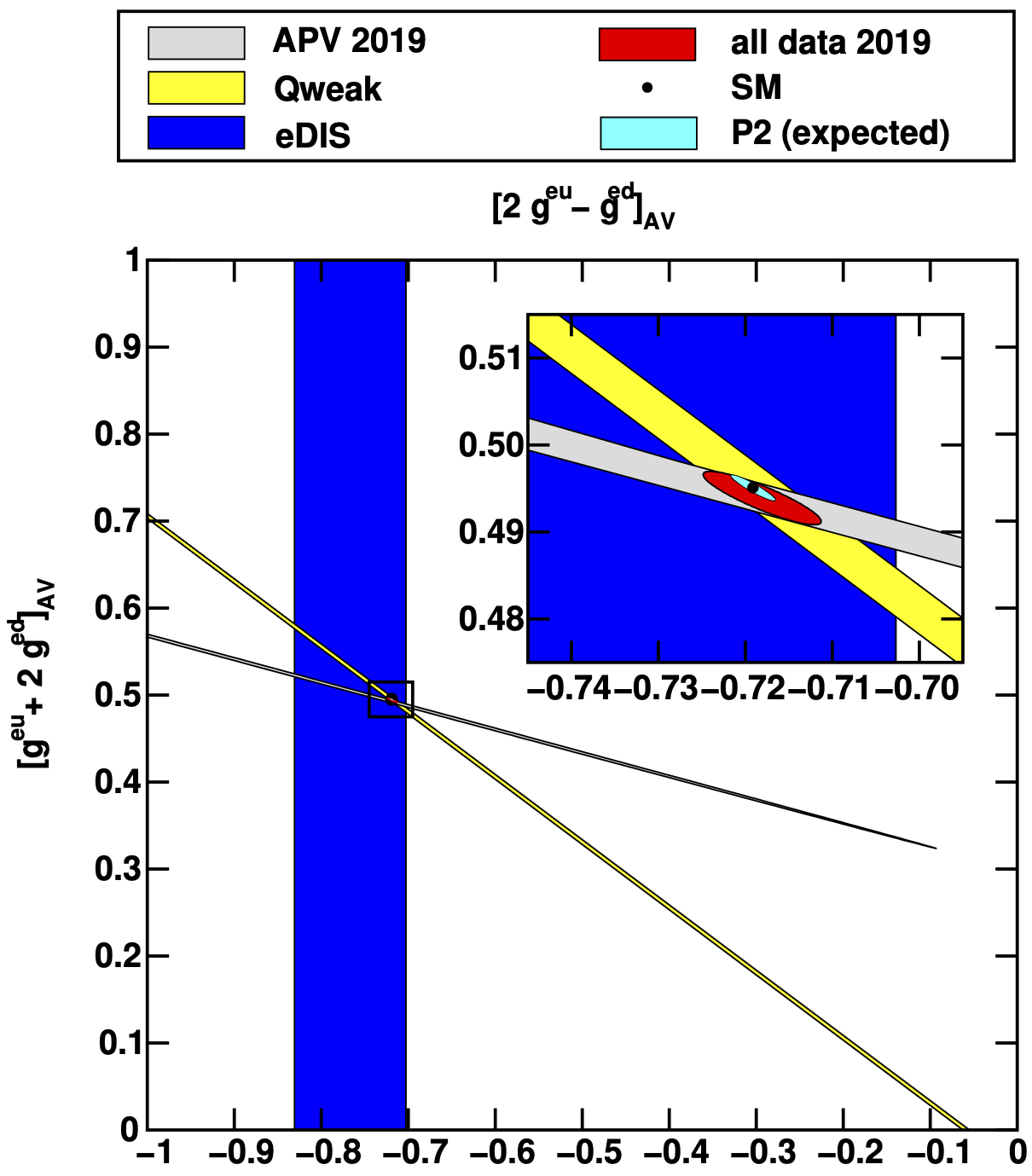}}
\end{minipage}
\hfill
\begin{minipage}{0.48\linewidth}
\centerline{\includegraphics[width=1.0\linewidth]{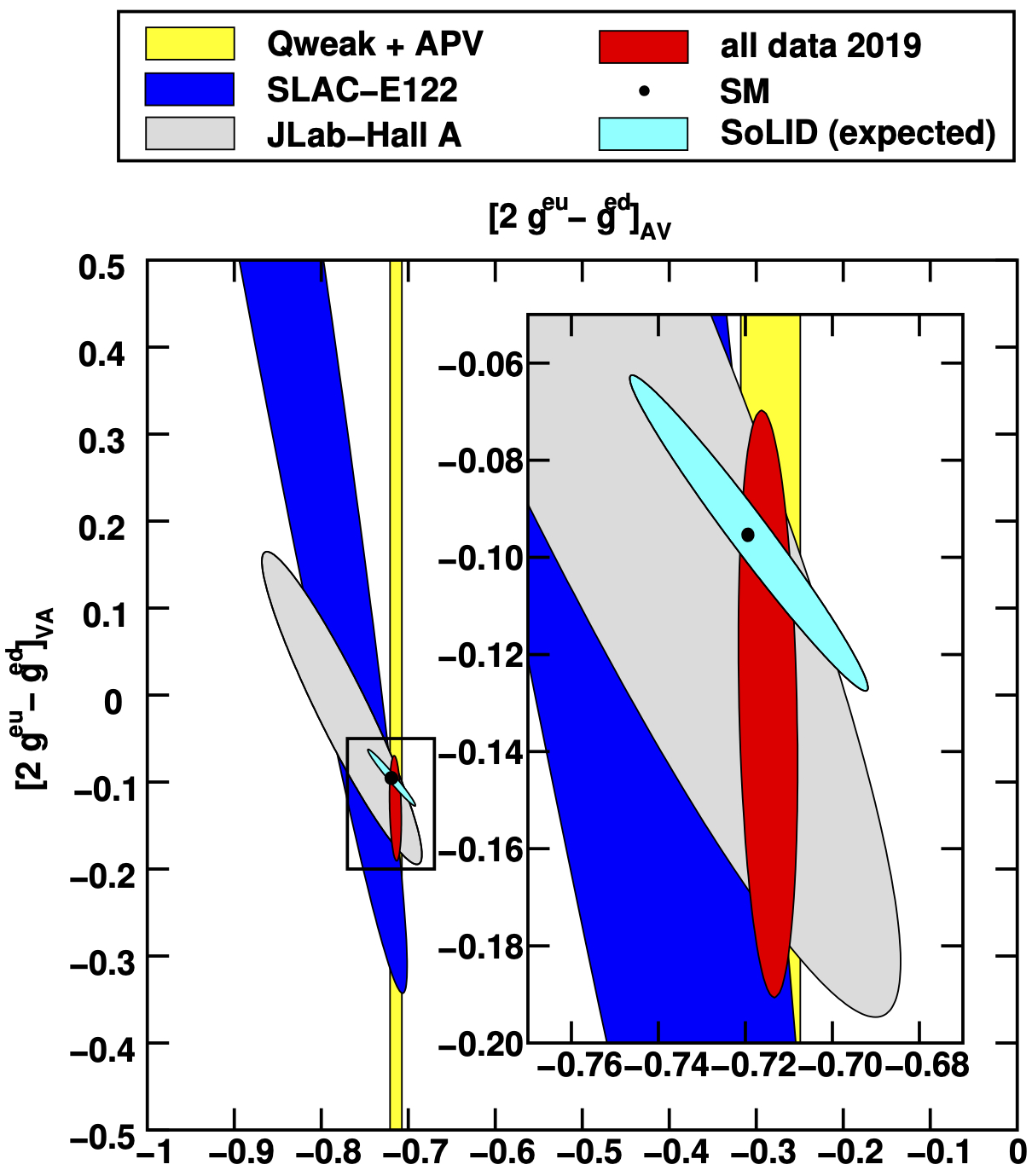}}
\end{minipage}
\caption[]{Results from APV and PVES.
Left: $C_1$ couplings.
Right: Electric charge-weighted $C_1$ and $C_2$ couplings.}
\label{fig:twins}
\end{figure}

Likewise, the 13 semi-leptonic operators include 4 scalar and 2 tensor structures, which do not interfere with photon or $Z$ exchange amplitudes.
The other 7 correspond to the vector- and axial-vector combinations, $e_V q_V$, $e_A q_V$, $e_V q_A$, and $e_A q_A$ (with $q = u$ or $d$),
which we will also call $C_0$, $C_1$, $C_2$, and $C_3$, respectively.  
Thus, it would seem there are 8 possibilities, but there is one constraint, 
$(\bar u_L \gamma^\mu u_L - \bar d_L \gamma^\mu d_L) \bar e_R \gamma_\mu e_R = 0$, 
because one operator is ruled out by $SU(2)_L$ gauge invariance.
The $C_1$ can be determined in APV and elastic PVES (Qweak and P2), 
while for the $C_2$ one needs to explore PVES in the deeply-inelastic regime (eDIS), 
such as in the experiment by the PVDIS Collaboration~\cite{Wang:2014bba} 
or with the future SoLID apparatus~\cite{Souder:2018xyr} at Jefferson Laboratory (JLab).
Unlike the $C_1$ and $C_2$, the $C_3$ are parity-even and may be studied by comparing $e^-$ and $e^+$ cross sections~\cite{Zheng:2021hcf}.
Current constraints on the $C_1$ and $C_2$ coefficients are shown in Fig.~\ref{fig:twins}.

\section{Some more recent results and conclusions}
\label{alphas}
The only determinations of the strong coupling, $\alpha_s$, that come with almost negligible QCD uncertainties are those from $Z$ pole measurements 
at LEP, most importantly from the $Z$ width, $\Gamma_Z$, the hadronic $Z$ peak cross section, $\sigma_{\rm had}^0$, 
and from the ratios of hadronic-to-leptonic $Z$ boson decay rates, $R_\ell$.
This is an old topic, but there was a very recent re-analysis~\cite{Janot:2019oyi} of the luminosity at LEP, partly
due to a new calculation of the small-angle Bhabha scattering cross section which is needed to measure it, and there were other refinements, as well.
As a result, the number of active neutrinos, $N_\nu$, extracted mostly from $\Gamma_Z$ and $\sigma_{\rm had}^0$ changed~\cite{Voutsinas:2019hwu} 
from what it was at the times of LEP.
But there is another consequence, which is that the combination,
\be
\alpha_s(M_Z) = 0.1228 \pm 0.0028,
\ee
of the EW $Z$ pole determinations is now roughly $2~\sigma$ higher than genuine QCD extractions. 
As a conclusion, the previous 2~$\sigma$ deficit in $N_\nu$ compared to the SM prediction $N_\nu = 3$ has migrated to $\alpha_s$.
An FCC-ee might be able to measure $\alpha_s(M_Z)$ to $\pm 5 \times 10^{-5}$ from these observables~\cite{Blondel:2019vdq}.

The $S$ and $T$ parameters correspond to the leading new physics contributions to the $W$ and $Z$ boson self-energies.
They present a nice and illustrative way to constrain BSM physics, where the right plot of Fig.~\ref{fig:weakplot} summarizes the current situation.
It is amusing to note that if one allows these two extra parameters, then the minimum $\chi^2$ of the global fit drops by 3.9~units,
which is quite unusual and perhaps this is telling us something, even though this is not very significant, yet.
Again, the FCC-ee would provide a tremendous improvement in precision, 
and the errors in $S$ and $T$ could drop by as much as a factor of twenty~\cite{Blondel:2019vdq}.

In conclusion, there is no conclusive evidence for new physics at the LHC so far, which is why there is a focus on the systematic 
SMEFT approach which is entirely model-independent, provided there are no new states with masses near the EW scale or below.
The recent LEP luminosity update confirmed that there are exactly 3~active neutrinos, but $\alpha_s$ is now slightly higher than expected.
Later in this decade we will witness many precise and complementary measurements of $\sin^2\theta_W$,
including ultra-high precision experiments at JLab and the MESA accelerator under construction in Mainz.
These will be competitive with the precision achieved by LEP and the SLC.
And in more future decades we hope to see next-generation lepton colliders like the ILC, the CEPC, the FCC-ee, CLIC, or a muon collider.

\section*{Acknowledgments}
This work is supported by the German-Mexican research collaboration grant SP 778/4–1 (DFG) and 278017 (CONACyT).
The hospitality and support of the Helmholtz Institute Mainz where part of the research presented here has been carried out 
is also gratefully acknowledged.


\section*{References}
\begin{thebibliography}{99}

\bibitem{Benesch:2014bas}
MOLLER Collaboration: J.~Benesch \textit{et al.},
arXiv:1411.4088 [nucl-ex].

\bibitem{Becker:2018ggl}
P2 Collaboration: D.~Becker \textit{et al.},
{\em Eur. Phys. J.} A \textbf{54}, 208 (2018).

\bibitem{PDG}
J.~Erler and A.~Freitas, {\sl Electroweak Model and Constraints on New Physics}, in Ref.~\cite{Zyla:2020zbs}.

\bibitem{Zyla:2020zbs}
Particle Data Group: P.~A.~Zyla \textit{et al.},
{\em PTEP} \textbf{2020}, 083C01 (2020).

\bibitem{Erler:2019hds}
J.~Erler and M.~Schott,
{\em Prog. Part. Nucl. Phys.} \textbf{106}, 68 (2019).

\bibitem{Blondel:2019vdq}
A.~Blondel \textit{et al.},
arXiv:1905.05078 [hep-ph].

\bibitem{Erler:2017knj}
J.~Erler and R.~Ferro-Hern\'andez,
{\em JHEP} \textbf{03}, 196 (2018).

\bibitem{Davoudiasl:2012qa}
H.~Davoudiasl, H.~S.~Lee and W.~J.~Marciano,
{\em Phys. Rev. Lett}. \textbf{109}, 031802 (2012).

\bibitem{Androic:2018kni}
Qweak Collaboration: D.~Androi\'c \textit{et al.},
{\em Nature} \textbf{557}, 207 (2018).

\bibitem{Henning:2015alf}
B.~Henning, X.~Lu, T.~Melia and H.~Murayama,
{\em JHEP} \textbf{08}, 016 (2017).

\bibitem{Anthony:2005pm}
SLAC--E158 Collaboration: P.~L.~Anthony \textit{et al.},
{\em Phys. Rev. Lett.} \textbf{95}, 081601 (2005).

\bibitem{Wang:2014bba}
PVDIS Collaboration: D.~Wang \textit{et al.},
{\em Nature} \textbf{506}, 67 (2014).

\bibitem{Souder:2018xyr}
P.~A.~Souder,
arXiv:1810.00989 [nucl-ex].

\bibitem{Zheng:2021hcf}
X.~Zheng, J.~Erler, Q.~Liu and H.~Spiesberger,
arXiv:2103.12555 [nucl-ex].

\bibitem{Janot:2019oyi}
P.~Janot and S.~Jadach,
{\em Phys. Lett.} B \textbf{803}, 135319 (2020). 

\bibitem{Voutsinas:2019hwu}
G.~Voutsinas, these proceedings.

\end{thebibliography}
\end{document}